\documentclass{elsart}

\usepackage{graphics}
\usepackage{graphicx}
\usepackage{amssymb}

\usepackage{bm}

\begin{document}
\begin{frontmatter}

\title{Basic properties of a vortex in a noncentrosymmetric superconductor
}

\author[AA]{N. Hayashi},
\author[CC]{Y. Kato},\,
\author[AA]{P. A. Frigeri},\,
\author[AA]{K. Wakabayashi},\,
\author[AA]{M. Sigrist}

\address[AA]{Institut f\"ur Theoretische Physik,
ETH-H\"onggerberg,
CH-8093 Z\"urich, Switzerland}  
\address[CC]{Department of Basic Science,
University of Tokyo,
Tokyo 153-8902, Japan}

\begin{abstract}
   We numerically study the vortex core structure
in a noncentrosymmetric superconductor such as CePt$_3$Si
without mirror symmetry about the $xy$ plane.
   A single vortex along the $z$ axis
and a mixed singlet-triplet Cooper pairing model
are considered.
  The spatial profiles of
the pair potential, local density of states, supercurrent density,
and radially-textured magnetic moment density around the vortex are obtained
in the clean limit
on the basis of the quasiclassical theory of superconductivity.
\end{abstract}

\end{frontmatter}

   Much attention has been focused on
the heavy fermion superconductor CePt$_{3}$Si,
which has
a noncentrosymmetric crystal structure
without mirror symmetry about the $xy$ plane \cite{bauer}.
   CePt$_{3}$Si is an extreme type-II superconductor
and
the vortex structure of the mixed state in this system
was recently studied
by Kaur {\it et al.} \cite{kaur}
and Yip \cite{yip}
on the basis of the Ginzburg-Landau theory and the London theory.
   In this paper,
we investigate the vortex core structure
on the basis of the quasiclassical theory of superconductivity \cite{serene},
which enables us to calculate more microscopically
the physical quantities such as
the pair potential, local density of states, supercurrent density,
and magnetic moment density.
   We consider a single vortex along the $z$ axis in the clean limit.

   The noncentrosymmetricity
(or the lack of inversion symmetry)
leads to the mixture of Cooper pairing channels
of different parity \cite{gorkov}.
 We consider the following superconducting order parameter
in a singlet-triplet mixing form:
$
{\hat \Delta}_k
=
\bigl(
  \Psi {\hat \sigma}_0
  + {\bm d}_k \cdot {\hat {\bm \sigma}}
\bigr)
i{\hat \sigma}_y
=
\Bigl[
\Psi {\hat \sigma}_0
 +
\Delta
\bigl(
  - {\tilde k}_y {\hat \sigma}_x
  + {\tilde k}_x {\hat \sigma}_y
\bigr)
\Bigr]
i{\hat \sigma}_y
$,
with the $s$-wave pairing component $\Psi$
and
the ${\bm d}$ vector
${\bm d}_k = \Delta (-{\tilde k}_y,{\tilde k}_x,0)$.
   This $s+p$-wave pairing state is proposed for CePt$_{3}$Si
in Ref.\ \cite{paolo3}.
Here,
$({\hat \sigma}_x,{\hat \sigma}_y,{\hat \sigma}_z)$
are the Pauli matrices in the spin space,
$ \hat{\sigma}_0 $ is the unit matrix,
and
${\tilde {\bm k}}
 =({\tilde k}_x,{\tilde k}_y,{\tilde k}_z)
 =(\cos\phi \sin\theta, \sin\phi \sin\theta, \cos\theta)$.

   The lack of inversion symmetry here
is incorporated through
a Rashba-type spin-orbit coupling with a form proposed in Ref.\ \cite{paolo1}.
   It splits the Fermi surface into two ones (I and II)
by lifting the spin degeneracy \cite{paolo3}.
   From the original Eilenberger equation
for noncentrosymmetric superconductivity \cite{haya1},
we obtain two equations corresponding to these split Fermi surfaces I and II
in the case of the above $s+p$-wave pairing state \cite{haya2},
\begin{equation}
i {\bm v}_{\rm I,II} \cdot
{\bm \nabla}{\check g}_{\rm I,II}
+ \bigl[ i\omega_n {\check \tau}_{3}
-{\check \Delta}_{\rm I,II},
{\check g}_{\rm I,II} \bigr]
=0,
\label{eq:eilen0}
\end{equation}
where
${\check \Delta}_{\rm I,II} =
\bigl[ ({\check \tau}_{1} + i {\check \tau}_{2}) \Delta_{\rm I,II}
- ({\check \tau}_{1} - i {\check \tau}_{2}) \Delta^{*}_{\rm I,II}
\bigr] /2$,
$\Delta_{\rm I,II}=\Psi \pm \Delta\sin\theta$
are the order parameters on the Fermi surfaces I and II,
$({\check \tau}_1,{\check \tau}_2,{\check \tau}_3)$
are the Pauli matrices in the particle-hole space,
and
the commutator
$[{\check a},{\check b}]={\check a}{\check b}-{\check b}{\check a}$.
   We neglect the vector potential in Eq.\ (\ref{eq:eilen0})
assuming the extreme type-II superconductivity.
   We use units in which $\hbar = k_{\rm B} = 1$.

   The Green functions ${\check g}_{\rm I,II}$
on the Fermi surfaces I and II are written
as a matrix in the particle-hole space,
\begin{equation}
{\check g}_{\rm I,II} ({\bm r}, {\tilde {\bm k}}, i\omega_n) =
-i\pi
\pmatrix{
g_{\rm I,II} &
if_{\rm I,II} \cr
-i{\bar f}_{\rm I,II} &
-g_{\rm I,II}
}.
\label{eq:qcg}
\end{equation}
   The regular Green function ${\hat g}$
as a matrix in the spin space is given
by \cite{paolo3,haya1,haya2}
\begin{eqnarray}
{\hat g}
&=&
g_{\rm I} {\hat \sigma}_{\rm I}
+
g_{\rm II} {\hat \sigma}_{\rm II}
\nonumber  \\
&=&
\frac{1}{2}
\pmatrix{
 g_{\rm I}+g_{\rm II} &
 -{\bar k}'_+ (g_{\rm I}-g_{\rm II})  \cr
 -{\bar k}'_- (g_{\rm I}-g_{\rm II})  &
 g_{\rm I}+g_{\rm II}
},
\label{eq:regular-g}
\end{eqnarray}
with
$  {\hat \sigma}_{\rm I,II} = (
{\hat \sigma}_0
   \pm {\bar {\bm g}}_k \cdot {\hat {\bm \sigma}}) /2 $
and 
$ {\bar {\bm g}}_k
= (-{\bar k}_y,{\bar k}_x,0) $.
 Here,
${\bar k}'_\pm = {\bar k}_y \pm i{\bar k}_x$
and
${\bar {\bm k}} = ({\bar k}_x,{\bar k}_y,0) = (\cos\phi,\sin\phi,0)$.

   We consider a single vortex which has a form,
$\Delta_{\rm I,II}(r,\phi_r;\theta)
=\bigl[
\Psi(r) \pm \Delta(r)\sin\theta
\bigr] \exp (i\phi_r)$.
   Here, the real-space coordinates
${\bm r}=(r\cos\phi_r,r\sin\phi_r,0)$,
and
the vortex center is situated at ${\bm r}=0$.
   The Fermi surface is assumed to be spherical,
and the differences of the density of states
and the Fermi velocity ${\bm v}_{\rm I,II}$
between the two Fermi surfaces I and II
are assumed to be small and are ignored.
   The results in this paper depend predominantly on
the spin structure [Eq.\ (\ref{eq:regular-g})] and
the gap structure on the 3D Fermi surfaces,
and the Fermi-surface anisotropy would not lead to 
qualitatively different results as long as the spin and gap topologies
are not altered.
   We numerically solve
the gap equations given in Ref.\ \cite{paolo3,haya1,haya2}
and the Eilenberger equations in Eq.\ (\ref{eq:eilen0})
self-consistently as in Ref.\ \cite{haya3}.
   When solving the gap equations,
we adopt the same values of parameters as used in Ref.\ \cite{haya1}.
   Thus, both the pair potentials $\Delta$ and $\Psi$
are real and positive, and $|\Delta|>|\Psi|$ \cite{haya1}.
   From now on, $T_{\mathrm c}$ is the superconducting critical temperature
and $\xi_0=v_{\mathrm F}/T_{\mathrm c}$ is the coherence length
at zero temperature
($v_{\mathrm F}=|{\bm v}_{\mathrm F}|$ is the Fermi velocity).

\begin{figure}[!ht]
\includegraphics[width=0.5\textwidth]{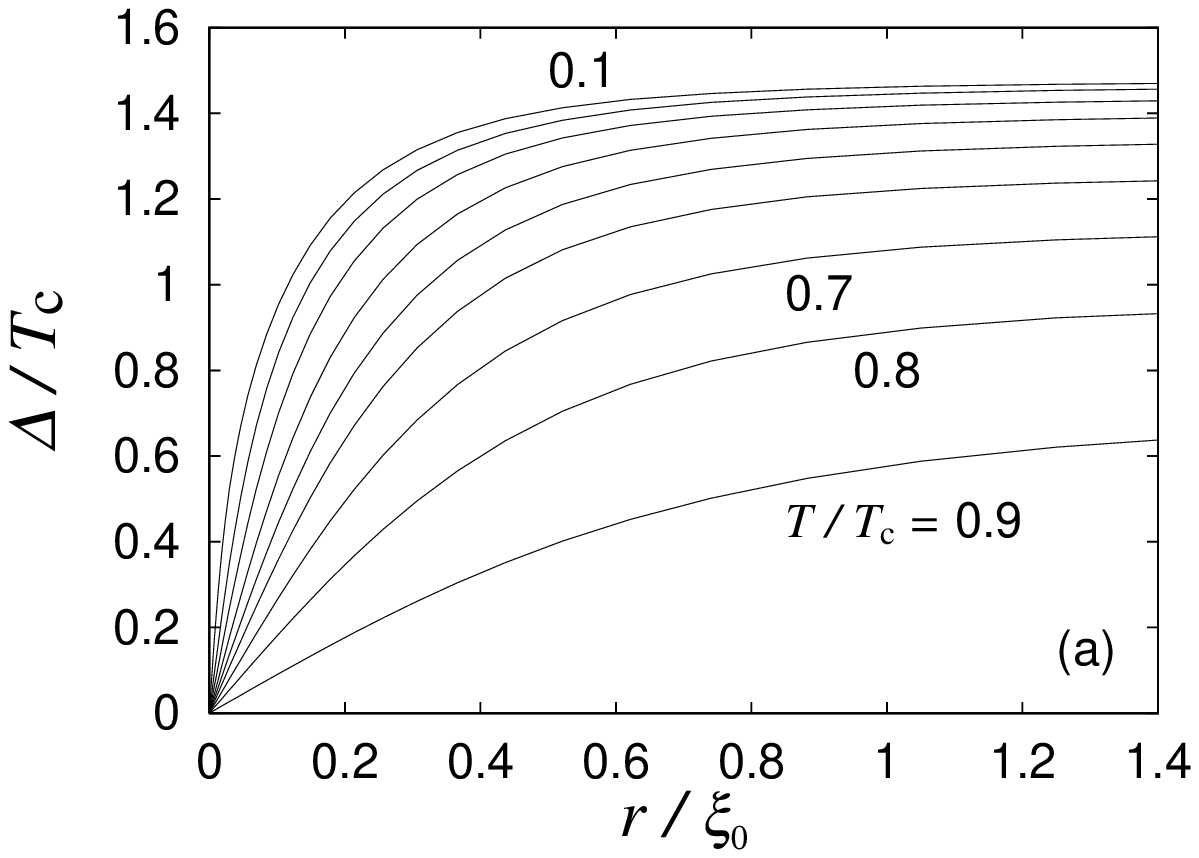}
\includegraphics[width=0.5\textwidth]{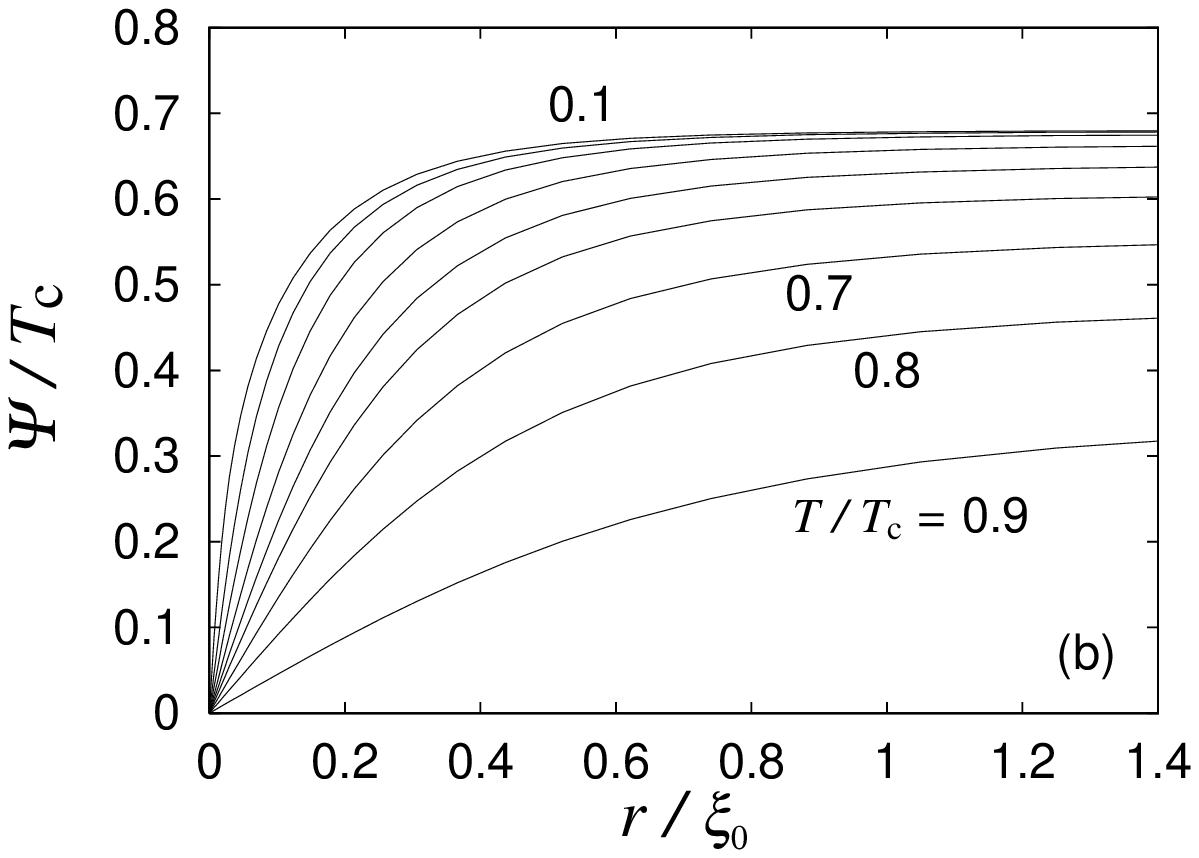}
\caption{
 Spatial profiles of the pair potentials.
 $T/T_{\mathrm c}=0.1$--0.9
from top to bottom by 0.1 step.
(a) the $p$-wave component $\Delta$, and (b) the $s$-wave one $\Psi$.
}
\label{fig1}
\end{figure}

   In Fig.\ \ref{fig1},
we show the spatial profiles of the pair potentials 
$\Delta$ ($p$-wave component)
and
$\Psi$ ($s$-wave one)
around the vortex for several temperatures $T$.
   It is noticed that
while the amplitude is different between $\Delta$ and $\Psi$,
the characteristic recovery length (namely, the core radius)
is the same for both.

\begin{figure}[!ht]
\begin{center}
\includegraphics[width=0.5\textwidth]{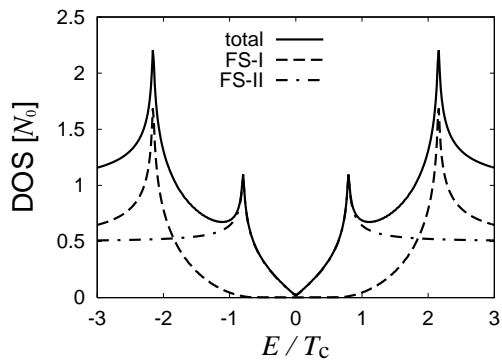}
\end{center}
\caption{
   The density of states in the bulk without vortices.
   $T/T_{\mathrm c}=0.1$
   and
   $\eta = 0.01 T_{\mathrm c}$.
}
\label{fig2}
\end{figure}

   The local density of states (per spin) is calculated by
\begin{eqnarray}
N(E,{\bm r})
&=&
\frac{N_0}{2} {\rm Re}
\Bigl\langle
   {\rm Tr}\bigl[
      {\hat g}(i\omega_n \rightarrow E +i\eta)
   \bigr]
\Bigr\rangle
\nonumber  \\
&=&
\frac{N_0}{2}
{\rm Re}
\Bigl\langle
      g_{\rm I}
      +
      g_{\rm II}
\Bigr\rangle,
\label{eq:LDOS}
\end{eqnarray}
where $\langle \cdots \rangle$ denotes
the average over the Fermi surface,
$N_0$ is the density of states per spin at the Fermi level,
and $\eta$ ($>0$) is the energy smearing factor.
   Before going into the vortex bound states,
let us see in Fig.\ \ref{fig2}
the density of states in the bulk without vortices.
   There are four gap edges (solid line).
   The system has two split Fermi surfaces I and II \cite{haya1,haya2}.
   The two of the gap edges
originate from the fully-gapped Fermi surface I (dashed line),
and
the other two originate from the line-node-gap Fermi surface II
(dash-dotted line).

\begin{figure}[!ht]
\begin{center}
\includegraphics[width=0.5\textwidth]{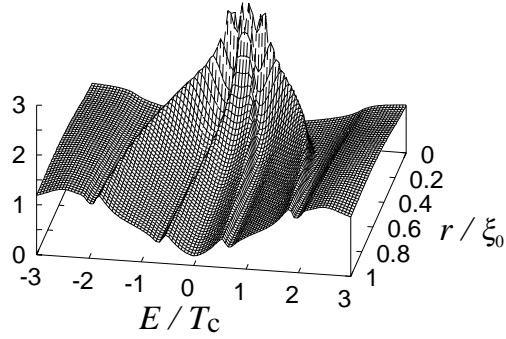}
\end{center}
\caption{
   The local density of states
$N(E,r)$
inside the vortex core in units of $N_0$.
   $T/T_{\mathrm c}=0.1$
   and
   $\eta = 0.05 T_{\mathrm c}$.
   Large zero-bias peak at $(E,r)=(0,0)$ is truncated in the figure.
}
\label{fig3}
\end{figure}

   In Fig.\ \ref{fig3},
we show the local density of states inside the vortex core.
   There are four branches of peaks,
which are related to the vortex bound states.
   The outer (inner) two branches originate from 
the vortex bound states of the quasiparticles on the Fermi surface I (II).
   Thus, the present spectra inside the vortex core in the clean limit
possess the same structure as those in a two-gap superconductor.

\begin{figure}[!ht]
\begin{center}
\includegraphics[width=0.5\textwidth]{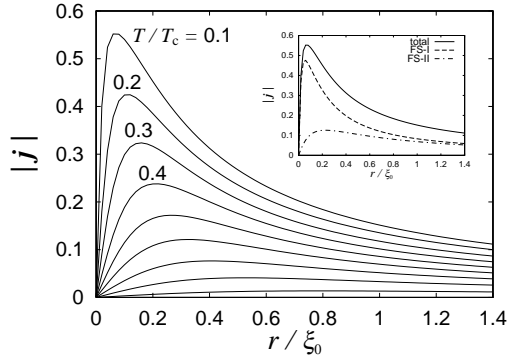}
\end{center}
\caption{
 Spatial profiles of the supercurrent density $|{\bm j}|$
in units of $2|e|v_{\mathrm F}N_0 T_{\mathrm c}$.
 $T/T_{\mathrm c}=0.1$--0.9
from top to bottom by 0.1 step.
 Inset: $|{\bm j}|$ at $T/T_{\mathrm c}=0.1$ (solid line),
the contribution of the Fermi surface I (dashed line),
and that of the Fermi surface II (dash-dotted line).
}
\label{fig4}
\end{figure}

   In Fig.\ \ref{fig4},
we plot the supercurrent density $|{\bm j}|$, which is calculated by
\begin{eqnarray}
{\bm j}
&=&
e T \sum_{\omega_n}
N_0
\Bigl\langle
   {\bm v}_{\mathrm F}
   {\rm Tr}\bigl[
      {\hat \sigma}_0
      (-i\pi {\hat g})
   \bigr]
\Bigr\rangle
\nonumber  \\
&=&
-i\pi e T \sum_{\omega_n}
N_0
\Bigl\langle
   {\bm v}_{\mathrm F}
      (g_{\rm I}
      +
      g_{\rm II})
\Bigr\rangle,
\label{eq:current}
\end{eqnarray}
where $e$ is the electric charge of the quasiparticle.
   We have confirmed numerically that
$|{\bm j}|$ decays as $\sim 1/r$ far away from the core.
   $|{\bm j}|$ exhibits essentially the same structure
as that in usual $s$-wave superconductors.

\begin{figure}[!ht]
\begin{center}
\includegraphics[width=0.5\textwidth]{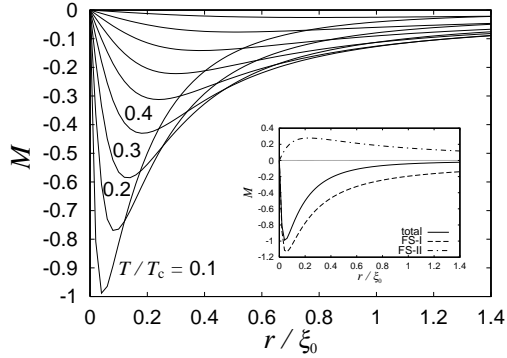}
\end{center}
\caption{
   The radial component of the magnetic moment density ${\bm M}$
around the vortex
in units of $\mu_{\mathrm B}N_0 T_{\mathrm c}$.
 $T/T_{\mathrm c}=0.1$--0.9
from bottom to top by 0.1 step.
   Inset: ${\bm M}$ at $T/T_{\mathrm c}=0.1$ (solid line),
the contribution of the Fermi surface I (dashed line),
and that of the Fermi surface II (dash-dotted line).
   The minus value means that
the sense of ${\bm M}$ is toward the vortex center.
}
\label{fig5}
\end{figure}

   Finally, we investigate
the magnetic moment density ${\bm M}$.
   The vortex-core magnetization
in the present noncentrosymmetric system has been reported
by Kaur {\it et al.} \cite{kaur} and Yip \cite{yip}.
   Here, we calculate it to obtain numeric results
at various temperatures
by means of a more microscopic derivation.
   ${\bm M}$ is calculated by
\begin{eqnarray}
{\bm M}
=
\mu_{\mathrm B} T \sum_{\omega_n}
N_0
\Bigl\langle
   {\rm Tr}\bigl[
      {\hat {\bm \sigma}}
      (-i\pi {\hat g})
   \bigr]
\Bigr\rangle,
\label{eq:mag}
\end{eqnarray}
where $\mu_{\mathrm B}$ is the magnetic moment of the quasiparticle.
   Substituting Eq.\ (\ref{eq:regular-g}) into this,
we obtain
\begin{eqnarray}
M_x
&=&
-i\pi
\mu_{\mathrm B} T \sum_{\omega_n}
N_0
\Bigl\langle
      (-{\bar k}_y)
      (g_{\rm I}-g_{\rm II})
\Bigr\rangle,
\label{eq:mag2}
   \\
M_y
&=&
-i\pi
\mu_{\mathrm B} T \sum_{\omega_n}
N_0
\Bigl\langle
      {\bar k}_x
      (g_{\rm I}-g_{\rm II})
\Bigr\rangle,
\label{eq:mag2-2}
   \\
M_z
&=&
0.
\end{eqnarray}
   We have confirmed numerically that
the azimuthal component of ${\bm M}$ is zero around the vortex,
namely ${\bm M}$ is aligned in the radial direction
in the $xy$ plane.
   In Fig.\ \ref{fig5},
we show the spatial profiles of the radial component of ${\bm M}$.
   We checked that $|{\bm M}|$ decays as $\sim 1/r$ far away from the core.
   The spatial profiles of $|{\bm M}|$ are similar to those of $|{\bm j}|$,
but there is a difference because of their microscopic origin.
   While $|{\bm j}|$ is composed of $g_{\rm I}+g_{\rm II}$
in Eq.\ (\ref{eq:current}),
$|{\bm M}|$ is composed of $g_{\rm I}-g_{\rm II}$
in Eqs.\ (\ref{eq:mag2}) and (\ref{eq:mag2-2})
[see also the insets of Figs. \ref{fig4} and \ref{fig5}].

   In conclusion,
we calculated the vortex core structure in the noncentrosymmetric
superconductor.
   We found that
the singlet and triplet components of the order parameter
have the same characteristic length (core radius) as each other.
   The local density of states
exhibits the two-gap property in the clean limit,
which is expected to be observed by STM \cite{STM1,STM2}
in order to clarify the validity of the present mixed-parity pairing model.
   We mention here that
impurity effects in the noncentrosymmetric
system can be different from those in simple two-gap systems in general,
therefore interesting phenomena in the density of states
are expected when the system deviates
from the clean limit.
   Detailed investigations
on such impurity effects are left for future studies.
   The radially-textured magnetic moment appears around the vortex,
which is expected to be observed directly
by spin-polarized STM \cite{spSTM}
in order to clarify
the electronic structure [Eq.\ (\ref{eq:regular-g})] specific to
the noncentrosymmetric superconductivity
with the rotating spin
on the Fermi surfaces as shown in a picture in Ref.\ \cite{saxena}.

   We thank D.~F.~Agterberg for stimulating discussion.
   We are grateful for financial support
from the Swiss Nationalfonds and the NCCR MaNEP.
   One of us (N.H.) is also supported by
2003 JSPS Postdoctoral Fellowships for Research Abroad.


\end{document}